\documentclass[%
 aps,
 pra,
 amsmath,
 amsfonts,
 amssymb,
 superscriptaddress,
 reprint
]{revtex4-1}

\usepackage{graphicx}
\usepackage{dcolumn}
\usepackage{bm}
\usepackage{textcomp}
\usepackage{siunitx}

\begin{document}

\title{Multistate metadynamics for automatic exploration of conical intersections}

\author{Joachim O. Lindner}
\affiliation{Institut f\"{u}r Physikalische und Theoretische Chemie, Julius-Maximilians-Universit\"{a}t W\"{u}rzburg,
Emil-Fischer-Str. 42, 97074 W\"{u}rzburg, Germany}

\author{Merle I. S. R\"{o}hr}
\affiliation{Institut f\"{u}r Physikalische und Theoretische Chemie, Julius-Maximilians-Universit\"{a}t W\"{u}rzburg,
Emil-Fischer-Str. 42, 97074 W\"{u}rzburg, Germany}
\affiliation{Center for Nanosystems Chemistry (CNC), Julius-Maximilians-Universit\"{a}t W\"{u}rzburg, Theodor-Boveri-Weg,
97074 W\"{u}rzburg, Germany}

\author{Roland Mitri\'{c}}
\email{roland.mitric@uni-wuerzburg.de}
\affiliation{Institut f\"{u}r Physikalische und Theoretische Chemie, Julius-Maximilians-Universit\"{a}t W\"{u}rzburg,
Emil-Fischer-Str. 42, 97074 W\"{u}rzburg, Germany}

\date{\today}
\begin{abstract}
We introduce multistate metadynamics for automatic exploration of conical intersection seams between adiabatic Born-Oppenheimer potential energy surfaces in molecular systems. By choosing the energy gap between the electronic states as a collective variable the metadynamics drives the system from an arbitrary ground-state configuration toward the intersection seam. Upon reaching the seam, the multistate electronic Hamiltonian is extended by introducing biasing potentials into the off-diagonal elements, and the molecular dynamics is continued on a modified potential energy surface obtained by diagonalization of the latter. The off-diagonal bias serves to locally open the energy gap and push the system to the next intersection point. In this way, the conical intersection energy landscape can be explored, identifying minimum energy crossing points and the barriers separating them. We illustrate the method on the example of furan, a prototype organic molecule exhibiting rich photophysics.
The multistate metadynamics reveals plateaus on the conical intersection energy landscape from which the minimum energy crossing points with characteristic geometries can be extracted. The method can be combined with the broad spectrum of electronic structure methods and represents a generally applicable tool for the exploration of photophysics and photochemistry in complex molecules and materials.

\end{abstract}
\maketitle

\section{Introduction}

\begin{figure*}
\begin{centering}
\includegraphics[width=0.85\textwidth]{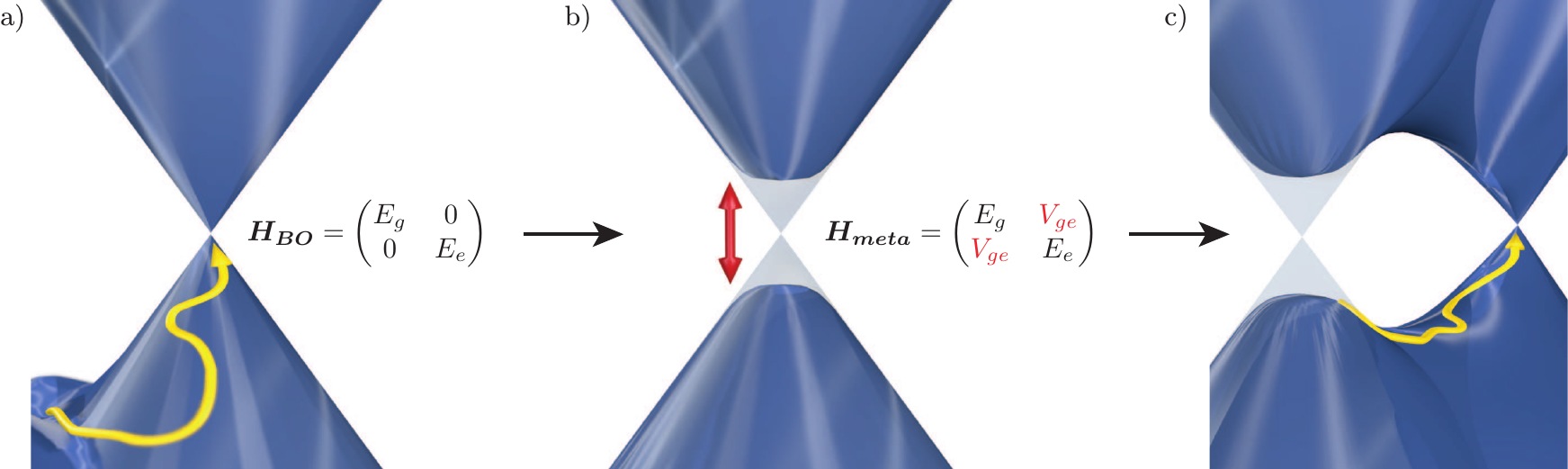}
\par\end{centering}
\protect\caption{Scheme of the algorithm. (a) Starting in the ground-state minimum, forces resulting from the bias
potential $V_{G}$ using energy gap CV drive the system toward the intersection seam (yellow trajectory). (b) Upon reaching the seam, an off-diagonal coupling $V_{ge}$ (red) is added to the electronic Hamiltonian, opening a gap in the vicinity of the intersection point. The metadynamics subsequently explores configurations with nearly degenerate ground- and excited-state energies until the next intersection point is reached (c). \label{fig:algorithm}}
\end{figure*}
The Born-Oppenheimer (BO) approximation, the cornerstone of molecular physics, breaks down
when molecules get electronically excited. This is most dramatically reflected in the presence of conical intersections (CI) \cite{Yarkony1996,Domcke2004-2011} between BO potential energy surfaces (PES).  A wide range of light-induced processes including vision in biology \cite{vbk1986,Polli2010}, photostability of DNA \cite{schultz2004}, a plethora of organic photochemical reactions \cite{olivucci1997}, and charge separation in photovoltaic devices \cite{Musser2015} are governed by the efficient nonradiative transitions mediated by a strong coupling between electronic and nuclear dynamics at CIs. In this sense, they play a role in photophysics and photochemistry analogous to the transition states in the ground-state chemistry. 
However, unlike the transition states, CIs are not isolated points on the 
PES but form a multidimensional seam. A significant effort has been invested both experimentally and theoretically to identify and characterize CIs \cite{mestdagh2003,mitric2006,Polli2010,kukura2014,mukamel2015,Mukamel2016}. A number of algorithms has been developed with the aim to search for the minimum energy configurations within the CI seam, since their knowledge allows us to predict the fate of a photoexcited molecule \cite{Manaa1993,Bearpark1994,Ciminelli2004}. In addition to these local optimization algorithms, methods have been also proposed for location of minimum energy crossing points (MECP), which are closest in terms of a mass-weighted distance to a predefined reference structure \cite{Levine2008}. 
However, all these methods require a reasonable initial guess for the CI geometry.
Alternatively, more elaborate approaches for the automated search of CI geometries based on the anharmonic downward distortion following (ADDF) 
and artificial force-induced reaction (AFIR) methods \cite{Maeda2009,Maeda2011a,Harabuchi2013,Maeda2015}, as well as the nudged elastic band method \cite{Mori2013}, have also been introduced and applied for the exploration of conical intersection seams.
Recent studies have emphasized the importance of conical intersections in coherent vibronic dynamics that involves nonadiabatic driving forces associated with the vibrational motion along the tuning modes \cite{Duan2016}. While such effects can only be properly investigated in the frame of quantum dynamics simulation, systematic and unbiased sampling of the conical intersection seam might help in construction of reliable (diabatic) Hamiltonians that are needed for such simulations.
Since CIs are usually found in the range of several eV above
the ground-state equilibrium structure, it would be desirable 
to develop an accelerated molecular dynamics technique that is able to reach and sample the CI seam starting directly from the ground-state minimum. 
The metadynamics, introduced by Parrinello and coworkers \cite{Laio2002,Barducci2008,Raiteri_2006,Bonomi2010,Tribello2010,Branduardi2012}, represents an ingeniously simple sampling method that uses collective variables (CVs) to drive transitions between different barrier-separated basins on the PES, allowing for systematic sampling of the PES as well as determination of free energies. Depending on the problem to be solved, a wide variety of CVs has been
used, ranging from simple geometrical quantities to complex variables
constructed by machine learning and dimensionality reduction techniques \cite{Tribello2010,Ceriotti2011}. 
In this article, we introduce a multistate metadynamics method for automatic exploration of conical intersection seams and identification of MECP.

\section{Method}
The central idea of our multistate metadynamics is to use the energy gap between the ground and excited electronic states as a collective variable to drive the system toward the CI starting from any ground-state structure. For this purpose, the system is propagated using the Newtonian equations of motion, which for the $i$th particle read
\begin{equation}
m_i \ddot{\mathbf{R}}_i=-\nabla _i[E_{g}+V_{G}(t)],\label{eq:mdforces}
\end{equation}
where $E_{g}$ represents the ground-state PES and $V_{G}(t)$ is a history-dependent bias potential:
\begin{equation}
\begin{aligned}
V_{G}(t)= &\sum_{t'=\tau_{G},2\tau_{G},\ldots}^{t}w\exp\left\{-\frac{\left[\Delta E_{meta}(t)-\Delta E_{meta}(t')\right]^{2}}{2\delta s^{2}}\right\} \\
&\times\Theta(\Delta E_{meta}(t')-\epsilon),\label{eq:metadynamics}
\end{aligned}
\end{equation}
which is dependent on the modified energy gap $\Delta E_{meta}$, and $w$ and $\delta s$ represent the fixed height and width, respectively. The bias potential is updated at regular time steps $\tau_G$ but only if the value of the gap is larger than a numerical threshold $\epsilon$, which is enforced by the presence of the Heaviside theta function $\Theta(\Delta E_{meta}-\epsilon)$ in Eq. (\ref{eq:metadynamics}). Starting from a minimum on the ground-state PES, the bias potential will drive the system toward the CI seam by systematically lowering the energy gap [see Fig. \ref{fig:algorithm}(a) for illustration]. 
After the intersection point is reached for the first time, the molecular electronic Hamiltonian, which is initially diagonal in the BO approximation, is extended by introducing a further biasing potential $V_{ge}$ into the off-diagonal elements according to
\begin{equation}
\bm{H_{BO}}=\begin{pmatrix}E_{g} & 0\\
0 & E_{e}
\end{pmatrix}
\rightarrow 
\bm{H_{meta}}=\begin{pmatrix}E_{g} & V_{ge}\\
V_{ge} & E_{e}
\end{pmatrix}.
\label{eq:hamilton_coupled}
\end{equation}
Subsequently, the metadynamics is continued on a locally modified potential energy surface obtained by diagonalization of the above Hamiltonian 
which gives rise to a modified PES with the effective energy gap
\begin{equation}
\Delta E_{meta}=\sqrt{\left(E_{g}-E_{e}\right)^{2}+4V_{ge}^{2}}.\label{eq:modified_gap}
\end{equation}
The off-diagonal bias $V_{ge}$ serves to locally open the gap between the eigenstates of the modified Hamiltonian. The metadynamics then drives the system to the next intersection point. In order to prevent the return to the previously sampled regions of the CI seam, the $V_{ge}$ is made dependent on a collective variable $s_{CI}$, which should be capable of distinguishing different molecular configurations and is updated only if the energy gap $\Delta E_{meta}$ is below $\epsilon$ according to
\begin{equation}
\begin{aligned}
V_{ge}(t)=&\sum_{t'=\tau_{G},2\tau_{G},\ldots}^{t}w\exp\left\{-\frac{\left[s_{CI}(t)-s_{CI}(t')\right]^{2}}{2\delta s^{2}}\right\}\\
&\times\Theta(\epsilon-\Delta E_{meta}(t')).\label{eq:metadynamics2}
\end{aligned}
\end{equation}
One possible choice of $s_{CI}$ is to use the distance matrix of a molecule, which is unique but inconvenient because of its high dimensionality and symmetry ambiguities. A better choice is to use its scalar invariants such as its lowest eigenvalue, or various topological indices \cite{Mihalic1992}. As a CV for the off-diagonal bias $s_{CI}$ in our simulations, we choose the 3D-Wiener number $W$ \cite{Bogdanov1989} defined as 
\begin{equation}
W=\frac{1}{2}\sum_{i}^{N}\sum_{j}^{N}d_{ij},\label{eq:wiener}
\end{equation}
where $N$ is the number of atoms and $d_{ij}$ are interatomic nonhydrogen distances. Its correlation with molecular shape made it a reliable topographical descriptor in numerous studies where structural unambiguity is important. Since the conical intersection seam is of dimension $f-2$, where $f$ is the number of degrees of freedom, and $W$ is a scalar function of the distances, there is no restriction to any specific region on the seam. Therefore, the most possible structural variability is achieved and can be further extended by running several trajectories with different initial conditions.

We wish to point out that as long as no Gaussians have been added to the off-diagonal bias $V_{ge}$, $\Delta E_{meta}$
exactly equals the BO gap $\Delta E_{BO}$. This can be derived from Eq.~(\ref{eq:modified_gap}) and is due to the construction of $\bm{H_{meta}}$ that reduces to $\bm{H_{BO}}$ in case of the nonexisting off-diagonal coupling.
Since $V_{ge}$ is updated only
in the vicinity of the CI seam, this holds true for the
complete pathway until the intersection is reached for the first
time [see Fig. \ref{fig:algorithm}(a)]. At this point, the metadynamics potential $V_{G}$ is
already overcompensating the minimum on the ground-state PES where
the simulation had been started, resulting in a persistent force toward
structures with small $\Delta E_{meta}$. If now a Gaussian bias is added to
$V_{ge}$ at the current value of the collective variable $s_{CI}$, $\Delta E_{meta}$ will increase, thus opening again the gap between the two states [see Fig.~\ref{fig:algorithm}(b)].
However, the bias potential $V_{G}$ disfavors nonzero values of the energy gap, which in turn forces the molecule to change the value of the collective variable $s_{CI}$ such that the intersection seam is reached again [see Fig. \ref{fig:algorithm}(c)].
In this way, the whole CI seam can be ``unzipped'', allowing for automatic exploration of its energy landscape.
In order to calculate the force in Eq.~(\ref{eq:mdforces}), the gradient of the modified energy gap is needed:
\begin{equation}
\color{red}
\nabla\left(\Delta E_{meta}\right)=\frac{\Delta E_{BO}\nabla\left(\Delta E_{BO}\right)+4V_{ge}\nabla\left(V_{ge}\right)}{\Delta E_{meta}},\label{eq:modified_gap_derivative}
\end{equation}
which requires the calculation of the gradients of the BO energies that can be provided by any suitable electronic structure method, as well as additional differentiation of $V_{ge}$:
\begin{equation}
\color{red}
\begin{aligned}
\nabla\left(V_{ge}\right)=&\sum_{t'=\tau_{G},2\tau_{G},\ldots}^{t}w\exp\left\{-\frac{\left[s_{CI}(t)-s_{CI}(t')\right]^{2}}{2\delta s^{2}}\right\}\\
&\times\Theta(\epsilon-\Delta E_{meta}(t'))\\
&\times\left\{-\frac{s_{CI}(t)-s_{CI}(t')}{\delta s^{2}}\nabla\left[s_{CI}(t)\right]\right\}.
\end{aligned}
\end{equation}
Like the potential $V_{ge}$ itself, it is given by a sum of Gaussians, but each of them is multiplied by an additional factor dependent on the gradient of the collective variable. For the 3D-Wiener number, $\nabla\left[s_{CI}(t)\right]$ is defined by
\begin{equation}
\nabla W=\frac{1}{2}\sum_i^N\sum_j^N\nabla d_{ij}.
\end{equation}
\begin{figure}
\begin{centering}
\includegraphics[width=0.45\textwidth]{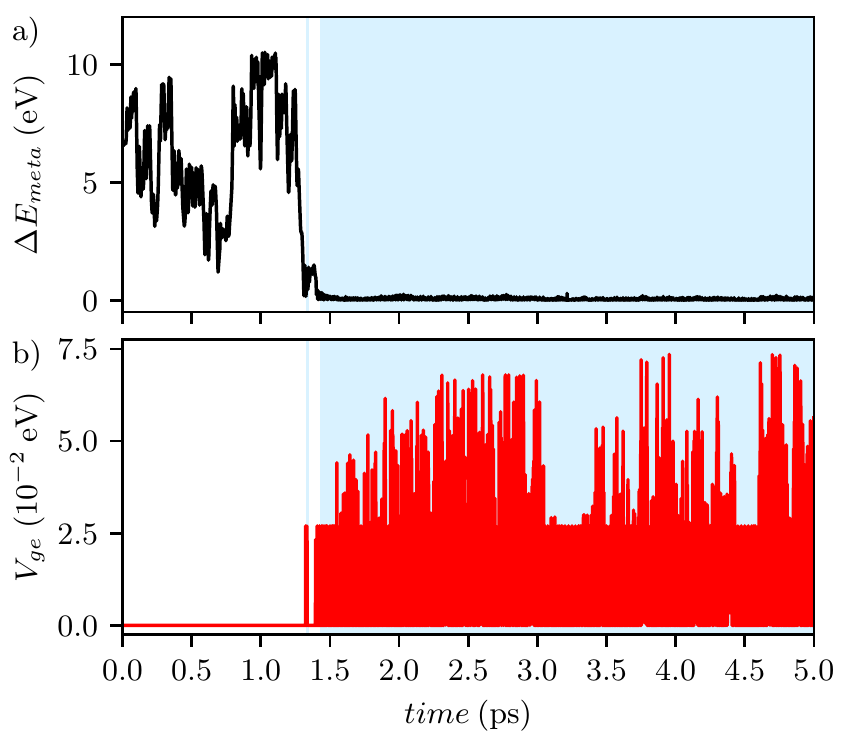}
\par\end{centering}
\protect\caption{Multistate metadynamics trajectory for furan:
(a) The energy gap $\Delta E_{meta}$ along the trajectory and (b) the off-diagonal bias potential $V_{ge}$. Periods in which $V_{ge}$ is updated ($\Delta E_{meta} < \epsilon$)
are highlighted in light blue.  \label{fig:trajectory-furan}}
\end{figure}
\begin{figure*}
\begin{centering}
\includegraphics{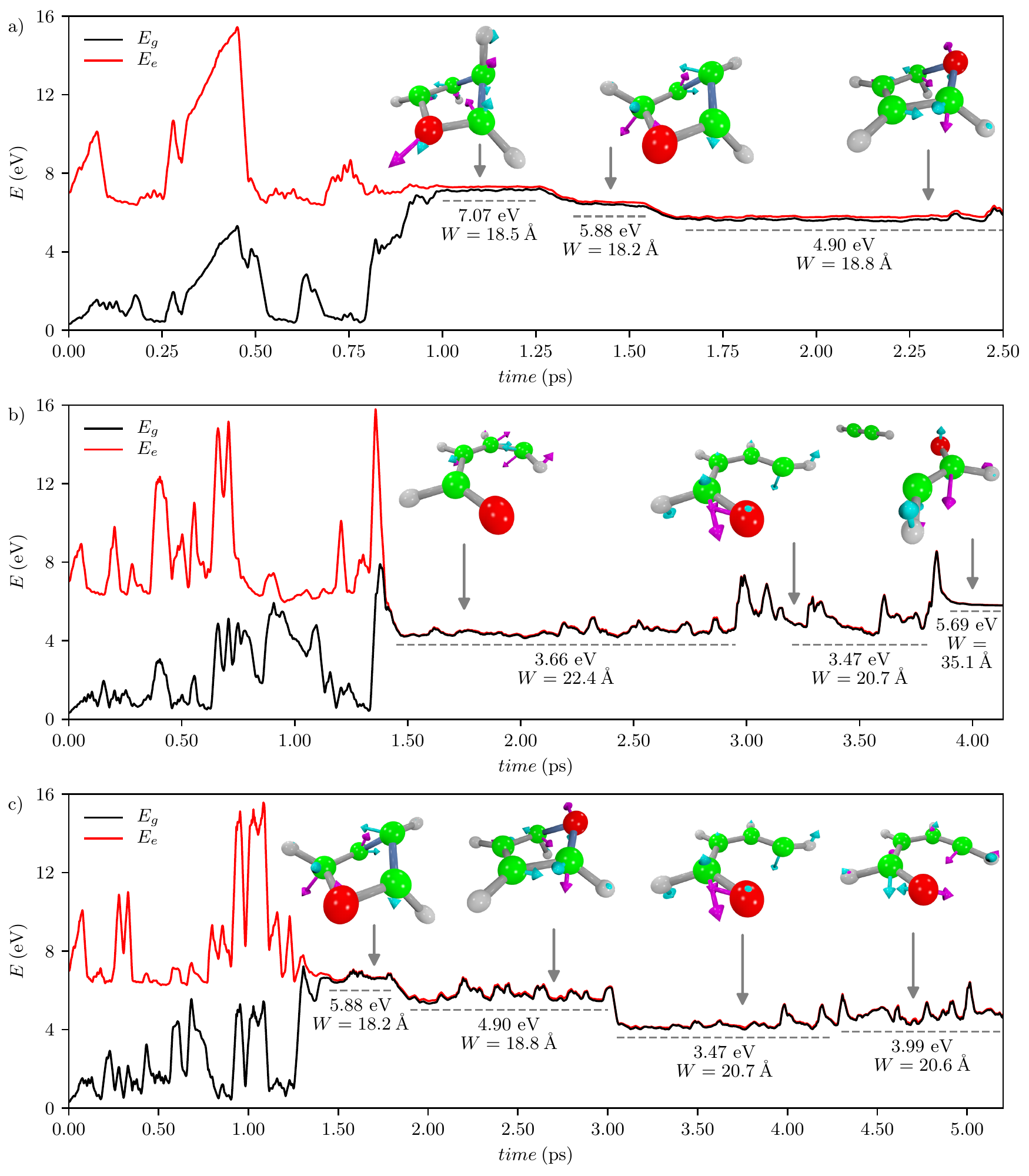}
\par\end{centering}
\protect\caption{Conical intersection energy landscapes of three different trajectories: Energies of the ground (black) and first excited (red) states along the  metadynamics trajectory for furan. A moving average with a window of \SI{25}{fs} has been applied for better visibility. Upon reaching the CI after 1.0--1.5 ps, the system remains confined to the CI seam and explores basins (dotted plateaus) corresponding to different MECPs. For each plateau, the optimized structures of the MECPs together with their branching space vectors, energies relative to the ground-state minimum (in eV), and values of the 3D-Wiener number are given. (a) Multiple ring puckering, (b) ring opening and fragmentation, and (c) combinations of both types are observed. \label{fig:opt-furan}}
\end{figure*}
The algorithm can be summarized with the following steps:
\begin{enumerate}
\item Initialize metadynamics trajectory with effective energy gap $\Delta E_{meta}$
as CV and choose an appropriate CV $s_{CI}$ for the off-diagonal bias potential.
\item Integrate the Newtonian equations of motion using a conventional molecular dynamics algorithm. If the actual time step is a multiple of $\tau_{G}$, proceed with one of the following steps:
\begin{enumerate}
\item If $\Delta E_{meta}>\epsilon$, add new Gaussian to the diagonal
metadynamics potential $V_{G}$.
\item If $\Delta E_{meta}<\epsilon$, add new Gaussian to the off-diagonal
coupling potential $V_{ge}$ and diagonalize the electronic Hamiltonian to obtain modified PES.
\end{enumerate}
\item Stop the simulation after a defined number of steps.
\item Identify the plateau regions in the CI energy landscape and select structures for MECP optimization.
\end{enumerate}
The use of a threshold $\epsilon$ in Eqs. (\ref{eq:metadynamics}) and (\ref{eq:metadynamics2})
for the switch between on- and off-diagonal metadynamics enables
the addition of Gaussians to $V_{ge}$ only if the trajectory is close
to the CI seam. At the same time, it prevents biasing
small values of $\Delta E_{meta}$ that would lead back to larger energy
gaps again. A reasonable choice for $\epsilon$ is to use the same
energy that is employed as Gaussian width for the $V_{ge}$.
The Gaussian width and height for the off-diagonal bias should be chosen sufficiently small in order to bias the previously visited configurations as local as possible. Thus, it can be ensured that the system stays in close proximity to the intersection seam most of the simulation time. Unlike conventional metadynamics, the aim of this algorithm is not to converge the metadynamics potential to the free energy surface and the results are therefore less sensitive to the choice of parameters.

Periods of the trajectory with low $\Delta E_{BO}$ represent structures close to the intersection seam. It is convenient to apply a local CI optimization method on snapshots from these regions in order to group the resulting MECPs by energy.

\section{Results and Discussion}

In order to illustrate the ability of the multistate metadynamics to sample conical intersections, we choose the furan molecule, which exhibits versatile photochemistry that is still not completely understood and has been a subject of a number of recent experimental and theoretical studies \cite{Stenrup2011,Gavrilov2008,Fuji2010,Spesyvtsev2015,Oesterling2017}.  
Here, the electronic structure of furan is described in the frame of the ab initio complete active space self-consistent field (CASSCF) method as implemented in the MOLPRO quantum chemical package \cite{Werner2011} with an active space of 10 electrons distributed over 9 orbitals together with the 6-31G{*} atomic basis set, as in previous quantum-chemical studies \cite{Oesterling2017}.
The Newtonian equations of motion in the metadynamics simulations were integrated using the velocity Verlet algorithm \cite{Verlet1967} with a step size of \SI{0.25}{fs} and the temperature was kept constant at \SI{300}{K} utilizing the Berendsen thermostat \cite{Berendsen_1984}.
Multistate metadynamics simulations were run using initial conditions sampled from a 2-ps-long MD trajectory propagated in the electronic ground state. Gaussian bias potentials with a width of
\SI{0.5}{eV} and a height of \SI{1.0}{eV} were added every 100 steps
(\SI{25}{fs}) to the diagonal bias. For the off-diagonal coupling $V_{ge}$, we apply the
3D-Wiener number (excluding hydrogen atoms) as CV with Gaussians of \SI{0.01}{\mathring{A}} width and \SI{0.027}{eV} height. The update
threshold $\epsilon$ was set to \SI{0.5}{eV}.
As can be seen from Fig.~\ref{fig:trajectory-furan}(a), the energy gap decreases during the multistate metadynamics simulation from an initial Franck-Condon value of \SI{6.97}{eV} to zero within the first 1.4 ps. At this point, the CI is reached for the first time and the off-diagonal coupling $V_{ge}$ shown in Fig.~\ref{fig:trajectory-furan}(b) is automatically switched on. Together with the previously built-up bias potential $V_G$, this forces the system to continue exploring the CI energy landscape.
The time evolution of the ground- and excited-state PES for three runs with different initial conditions sampled from the ground state trajectory is shown in Fig.~\ref{fig:opt-furan}. In each case, the system is initially driven from a local minimum toward the conical intersection, leading to the closure of the energy gap. Subsequently, the dynamics explores the CI landscape and the energy gap remains zero within the predefined numerical threshold. As can be seen from Fig. ~\ref{fig:opt-furan}, CI energy landscape exhibits several plateaus and depending on the CI character, some of them are separated by relatively large barriers. Low barriers are in general present for transitions between various ring-puckering CIs, as is seen in the trajectory depicted in Fig.~\ref{fig:opt-furan}(a). On the contrary, the trajectory shown in Fig.~\ref{fig:opt-furan}(b) leads to the CI structures exhibiting ring opening as well as fragmentation. Molecular fragmentations are typically found after longer simulation times since they are characterized by large Wiener numbers. It is interesting to note that both molecular geometries that have previously been held responsible for the ultrafast deactivation of furan \cite{Fuji2010, Oesterling2017}, a ring puckering of the oxygen atom and the low-energetic $\mathrm{C-O}$ ring opening, have been identified in successive order in the third trajectory shown in Fig.~\ref{fig:opt-furan}(c).
The found geometries have been subsequently fully optimized using the Bearpark-Robb local optimization algorithm \cite{Bearpark1994} and their branching plane vectors have been determined (cf. Fig.~\ref{fig:opt-furan}). 

\section{Conclusion}
In summary, we have developed a multistate extension of the metadynamics with the aim to automatically explore conical intersection seams between BO PES.  Biasing metadynamics potentials are introduced as off-diagonal elements into the multistate electronic Hamiltonian and MD simulations are run on modified potential energy surface obtained by diagonalization of the latter, using the energy gap as a collective variable. The method can be easily implemented in the frame of any electronic structure method capable of providing energy gradients and excitation energies. It can be applied to explore photochemical reaction pathways, nonradiative relaxation channels, and photophysics of complex molecular systems. As an illustration, we have performed simulations starting from the ground-state structure of the furan and have demonstrated that the conical intersection landscape can be efficiently mapped, allowing us to systematically identify a large number of minimum energy crossing points that can mediate nonradiative relaxation and photochemical reactivity.

\section*{Acknowledgement}
Funding by the European Research Council (ERC) Consolidator Grant DYNAMO (Grant No. 646737)
is gratefully acknowledged.
\bibliographystyle{apsrev4-1}
\bibliography{references}

\end{document}